\documentclass[twocolumn,citeautoscript,pra,superscriptaddress,amsmath,amssymb,8pt]{revtex4-1}
\usepackage{graphicx}
\usepackage{gensymb}
\usepackage{color}
\usepackage[dvipsnames]{xcolor}
\usepackage{float}
\usepackage{upgreek}
\usepackage{amsmath,amssymb}

\graphicspath{{C:/Users/Alex/Desktop/plots/shallow_hpht/}}

\begin{document}

\title{On the creation of near-surface nitrogen-vacancy centre ensembles by implantation of type Ib diamond}

\author{A. J. Healey}
\email{alexander.healey@unimelb.edu.au}
\affiliation{School of Physics, University of Melbourne, VIC 3010, Australia}
\affiliation{Centre for Quantum Computation and Communication Technology, School of Physics, University of Melbourne, VIC 3010, Australia}	

\author{S. C. Scholten}
\affiliation{School of Physics, University of Melbourne, VIC 3010, Australia}
\affiliation{Centre for Quantum Computation and Communication Technology, School of Physics, University of Melbourne, VIC 3010, Australia}	

\author{A. Nadarajah}
\affiliation{School of Physics, University of Melbourne, VIC 3010, Australia}

\author{Priya Singh}
\affiliation{School of Science, RMIT University, Melbourne, VIC 3001, Australia}

\author{N. Dontschuk}
\affiliation{School of Physics, University of Melbourne, VIC 3010, Australia}

\author{L. C. L. Hollenberg}
\affiliation{School of Physics, University of Melbourne, VIC 3010, Australia}
\affiliation{Centre for Quantum Computation and Communication Technology, School of Physics, University of Melbourne, VIC 3010, Australia}	

\author{D. A. Simpson}
\affiliation{School of Physics, University of Melbourne, VIC 3010, Australia}

\author{J.-P. Tetienne}
\affiliation{School of Science, RMIT University, Melbourne, VIC 3001, Australia}

\begin{abstract}
Dense, near-surface (within $\sim 10$~nm) ensembles of nitrogen-vacancy (NV) centres in diamond are rapidly moving into prominence as the workhorse of a variety of envisaged applications, ranging from the imaging of fast-fluctuating magnetic signals to the facilitation of nuclear hyperpolarisation. Unlike their bulk counterparts, near-surface ensembles suffer from charge stability issues and reduced NV formation efficiency due to the diamond surface's role as a vacancy sink during annealing and an electron sink afterwards. To this end, work is ongoing to determine the best methods for producing high-quality ensembles in this regime. Here we examine the prospects for creating such ensembles cost-effectively by implanting nitrogen-rich type Ib diamond with electron donors, aiming to exploit the high bulk nitrogen density to combat surface-induced band bending in the process. This approach has previously been successful at creating deeper ensembles, however we find that in the near-surface regime there are fewer benefits over nitrogen implantation into pure diamond substrates. Our results suggest that control over diamond surface termination during annealing is key to successfully creating high-yield near-surface NV ensembles generally, and implantation into type Ib diamond may be worth revisiting once that has been accomplished. 
\end{abstract}

\maketitle 
\section{Introduction}
Shallow nitrogen-vacancy (NV) centres in diamond have been shown to be useful as sensors of weak fluctuating magnetic signals \cite{Cole2009,Meriles2010,Staudacher2013,Mamin2013,Loretz2014,Muller2014,Devience2015, Lovchinsky2016,Shi2015} and as a potential vehicle for enabling hyperpolarisation of nuclear spins external to the diamond \cite{Shagieva2018,Fernandez-Acebal2018,Broadway2018a,Tetienne2021}. Much work to-date has focussed on the use (and production) of near-surface single NVs that can sometimes exist in the required negative charge state within a few nanometres of the surface in spite of the unfavourable local Fermi level position~\cite{Collins2002}. Increasingly, however, applications such as scaled-up hyperpolarisation~\cite{Healey2021,Rizzato2022} and imaging of AC magnetic fields~\cite{Steinert2013,Devience2015,Simpson2017,McCoey2020,Ziem2019,Bertelli2020} demand high-density ensembles of stable near-surface NVs. When sampling a large number of NVs, the impact of surface-induced band-bending becomes clear with the NV$^-$ depth distribution cutting off at 6-7~nm from the surface~\cite{Broadway2018b,Healey2021}. Taken in combination with the expectation that vacancies produced by near-surface implants, required to form NV centres during a subsequent annealing process, will tend to out-diffuse to the surface~\cite{Pezzagna2010,Racke2021}, NV$^-$ yields in such ensembles are much lower than their bulk counterparts. 

In the bulk-like regime (mean ensemble depth $d_{\rm NV}$ order 100~nm or more), where the creation of high-density NV ensembles is comparatively well-developed, it has been shown that starting with N-rich (type Ib) diamond grown via the high-pressure high-temperature (HPHT) method and implanting with arbitrary ions is successful in producing a well-localised sensing layer with quantum properties that compete well with other methods based on high-quality chemical vapour deposition (CVD) growth~\cite{Huang2013,Mccloskey2014,Fescenko2019,Healey2020}. Extending these results to the near-surface regime is attractive due to the relative cost-efficiency and accessibility of this technique. Additionally, one may wonder whether the high nitrogen density of the bulk crystal is effective in combating the near-surface band-bending. However, the issue of vacancy diffusion during annealing looms as an impediment to high-yield ensemble formation: under typical annealing temperatures (800-900$\degree$C) various studies have shown that the vacancy diffusion length may extend as high as 300~nm~\cite{Orwa2012,Alsid2019,Racke2021}. Using a random walk model, R\"acke \textit{et al.}~\cite{Racke2021} showed that the reduced near-surface NV yield typically observed is largely explained by vacancy out-diffusion, even without taking the surface to be a vacancy attractor. Additionally, diffusion \textit{into} the diamond is also problematic as the envisaged applications require that NVs be confined within $\approx 10$~nm of the surface, although in this case we would hope that the actual diffusion length would fall well short of the theoretical upper bound due to substitutional N acting as efficient vacancy sinks.

In this study, we examine the merits of creating near-surface NV ensembles through ion implantation of commercially-sourced type Ib HPHT diamond, in view of the factors outlined above. By implanting diamonds containing distinct growth sectors (each with a characteristic native N density), we are able to control for the effect of the bulk N density to determine the role this plays on ensemble surface proximity and yield. We also implant at multiple depths (set by the implant energies) and with two levels of vacancy production (given by the implantation dose and atomic species) to assess the practical role of vacancy diffusion in the high-N regime. The quality of the ensembles produced is assessed by making measurements of the NV yield and their quantum coherence. We conclude with a discussion of the limitations of the study and the prospects for future work. 

\section{Experimental section}
\subsection{Diamond preparation}
A series of type Ib HPHT diamond substrates (purchased from Delaware Diamond Knives) containing sectors with varying levels of native nitrogen were subjected to ion implantation processes (InnovION) to form NV ensembles. To control for as many variables as possible and ensure comparisons between sectors are valid, only two different diamonds were used (initial size 4$\times 4\times 0.1$~mm). These diamonds were then laser cut into smaller pieces to undergo different preparation. The implant parameters were chosen to create vacancy profiles peaking at approximately 3, 4, and 5~nm from the diamond surface. The vacancy profiles were predicted using stopping range in matter (SRIM) simulations, shown in Fig.~\ref{fig1}(a). Neglecting charge state and vacancy diffusion considerations, we expect to produce a uniform NV layer of width $w_{\rm SRIM}$, which extends from the surface to the depth where the vacancy production decreases below 50~ppm [dotted line in Fig.~\ref{fig1}(a)], an approximate NV creation saturation threshold previously identified in the bulk regime~\cite{Healey2020}. SRIM simulations assume an amorphous substrate so we implanted our samples with a sample tilt of 7$\degree$ to minimise ion channelling. The first set of implants chosen to meet this criteria were $^{16}$O at a dose of 1$\times 10^{12}$~cm$^{-2}$ at energies of 2.5, 4, and 6~keV respectively. A second set of implants designed to produce an order of magnitude more vacancies with similar depth profiles were $^{31}$P implants at a dose of 5$\times 10^{12}$~cm$^{-2}$ with energies of 4, 7, and 11~keV respectively. In both cases the implant species were chosen to be electron donors to the diamond lattice in an attempt to further offset the band bending from the surface. 

For the first set of implants we also controlled for the surface preparation. The as-purchased diamonds arrived with a polished surface finish (Ra $<5$~nm) and an oxygen reactive ion etching (RIE) process can be used to remove polishing damage. For the O implants we only performed RIE on some of the substrates to see if the process made a difference to NV yield or quantum properties. Following implantation, all samples were annealed in a vacuum furnace (pressure held below $10^{-5}$~hPa) using a ramp sequence that culminated with one hour at 800$\degree$C (2~h ramp to 400$\degree$C, 3~h at 400$\degree$C, 3~h ramp to 800$\degree$C, 1~h at 800$\degree$C, 2~h ramp to room temperature). The one hour plateau was chosen in an attempt to maximise NV yield while minimising vacancy diffusion into the diamond, in practice there is expected to be a trade-off between these two factors. The diamonds were then cleaned in a boiling mixture of sulphuric and nitric acid to achieve a standardised, oxygen-terminated surface. 

\subsection{NV yield}
To determine the NV yield in our samples, we used a confocal microscope to measure the photoluminescence (PL) count rate per unit area (filtering with a 660-735~nm band pass filter) and translated this to an areal NV density $\sigma_{\rm NV}$ by dividing by the PL given by a single NV centre under the same excitation and collection conditions. We can then consider two yield metrics: the conversion of native nitrogen and created vacancies to NV centres (dubbed N-to-NV and V-to-NV yields respectively). The N-to-NV yield is given by the ratio [NV]/[N], where [NV]$=\sigma_{\rm NV} / w_{\rm SRIM}$ and [N] is the native N density of a given growth sector. [N] was deduced by measuring the Hahn echo $T_2$ and taking the relationship determined by Bauch \textit{et al.}~\cite{Bauch2020}, where the $T_2$ was measured away from the influence of the surface where possible. We note that [N] could be overestimated if the nitrogen bath is not the dominant source of decoherence (most relevant for the less dense sectors) and that $\sigma_{\rm NV}$ could be overestimated by the presence of background fluorescence or PL due to the neutral NV charge state. 

\begin{figure}
\centering
\includegraphics[width=0.45\textwidth]{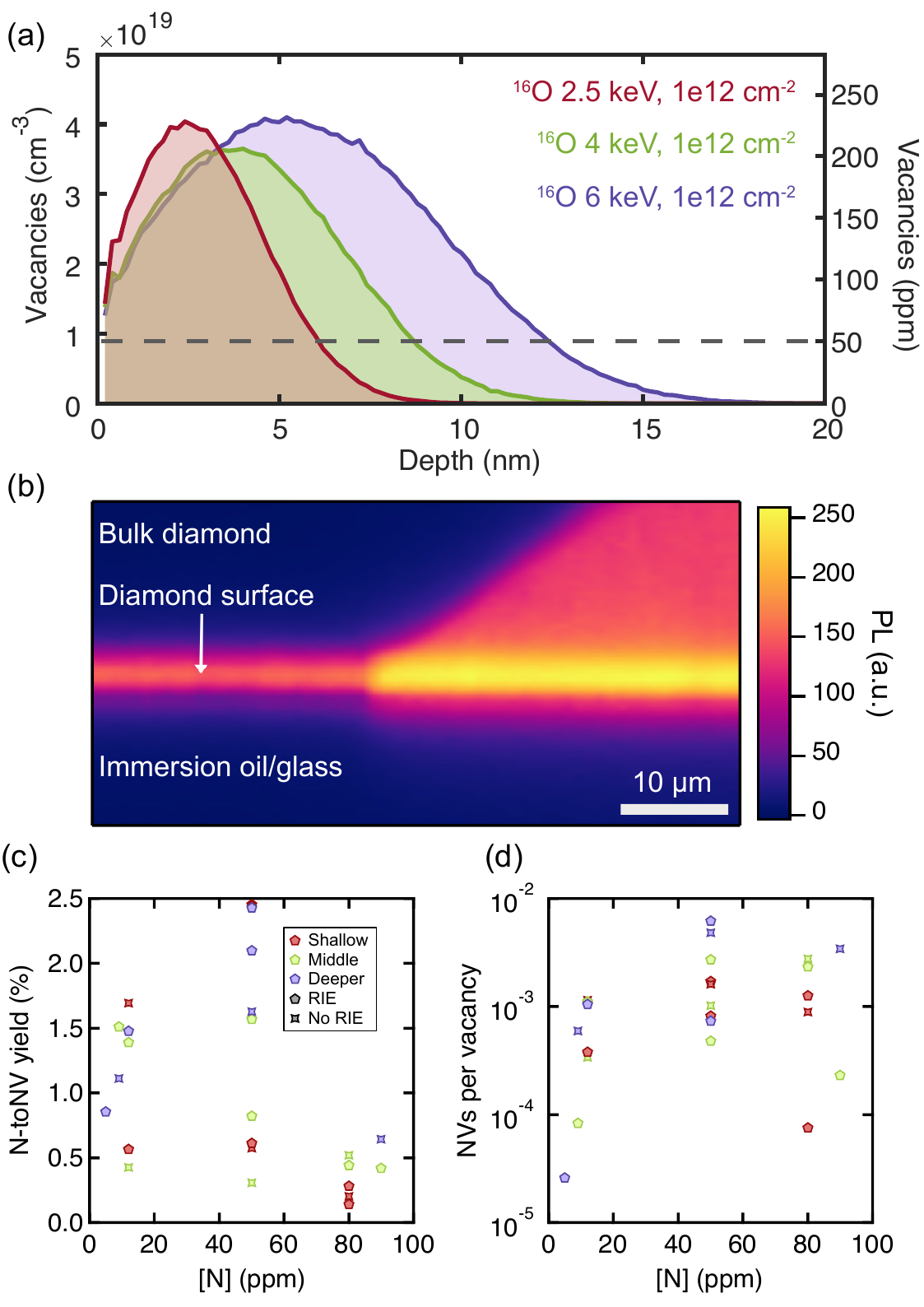}
\caption{\textbf{Creating shallow NV layers in type Ib diamond} (a) SRIM simulations of the oxygen implants conducted, taking a 7$\degree$ angle of incidence. Phosphorus implants were also conducted at energies to approximately match the expected vacancy depth profile, but creating an order of magnitude more vacancies. (b) Confocal $xz$ scan of one diamond sample, showing an NV layer localised at the surface. Two sectors are visible, with the left and right hand regions' nitrogen content estimated at 8 and 50~ppm respectively. (c) NV yield (N-to-NV) estimated as described in the text. Colouring shown in the legend matches the implants depicted in (a). (d) Plot of NVs created per vacancy, taking the NV yield as in (c) and comparing against the total vacancy production predicted by SRIM, again plotted versus nitrogen concentration.}
\label{fig1}
\end{figure}

An example $xz$ confocal scan is shown in Fig.~\ref{fig1}(b). A well-defined NV layer is present at the diamond surface although the resolution of the scan is not high enough to determine if the layer's extent matches the vacancy distribution predicted by SRIM. In this image we can see two growth sectors containing variable amounts of nitrogen: the right hand sector (estimated nitrogen density [N]=50~ppm compared to 8~ppm for the left hand sector) has significant background PL away from the surface and the PL of the near-surface sensing layer also varies with the native nitrogen density. In both cases, however, the near-surface sensing layer PL greatly exceeds the background for a given sector, indicating locally increased NV conversion as expected. 

Figure~\ref{fig1}(c) shows the computed N-to-NV yields plotted against the inferred native N density of a given diamond sector, with the marker colouring indicating the implant energies. The highest yields are close to 2.5\%, however a majority of regions have yields of less than 1\%, particularly for higher-N sectors and shallower implants. As we filter the PL for the negatively-charged NV centre, these yields are not necessarily reflective of the total NV conversion but rather conversion to the charge state useful for sensing and hyperpolarisation applications. The reduced yields compared to deeper implants~\cite{Healey2020} therefore could be band-bending-induced or due to reduced creation efficiency independent of charge state. The yields in these samples are comparable to typical N-implants~\cite{Healey2021} but do not appear to offer an advantage in general. 

Looking at the V-to-N yield (``vacancy yield"), plotted in Fig.~\ref{fig1}(d), may give a clue as to the origin of the poor conversion. Taking the vacancy creation predicted by SRIM for each implant, we find that around $10^{-3}$ NVs are created per vacancy implanted in most cases, consistent with the modelling of R\"acke \textit{et al.}~\cite{Racke2021} for the case of the diamond surface acting as a vacancy sink. The spread in vacancy yield may be due to variable surface termination, motivating further study into maintaining high-quality surface termination during annealing so as to keep more vacancies within the diamond. No obvious trends were present within our data based on the two surface preparations carried out, however. 

The lacklustre vacancy conversion observed for the oxygen implants motivated additional implants to be carried out, using $5\times 10^{13}$~cm$^{-3}$ $^{31}$P implants at energies designed to match the vacancy production profile of the oxygen implants. These phosphorus implants are expected to have produced an order of magnitude more vacancies, however we find that the N-to-NV yield is not improved, meaning that the useful vacancy creation threshold identified in previous work~\cite{Healey2020} of around 50~ppm appears to be retained in this near-surface regime, despite overall lower NV creation efficiency. The interpretation may be that in this high vacancy production regime, the formation of multi-vacancy clusters is more predominant, which either anneal out or add a source of spin noise~\cite{Tetienne2018}, and therefore the number of vacancies available to form NVs is not much greater. 
\subsection{NV depth}
\begin{figure}
\centering
\includegraphics[width=0.45\textwidth]{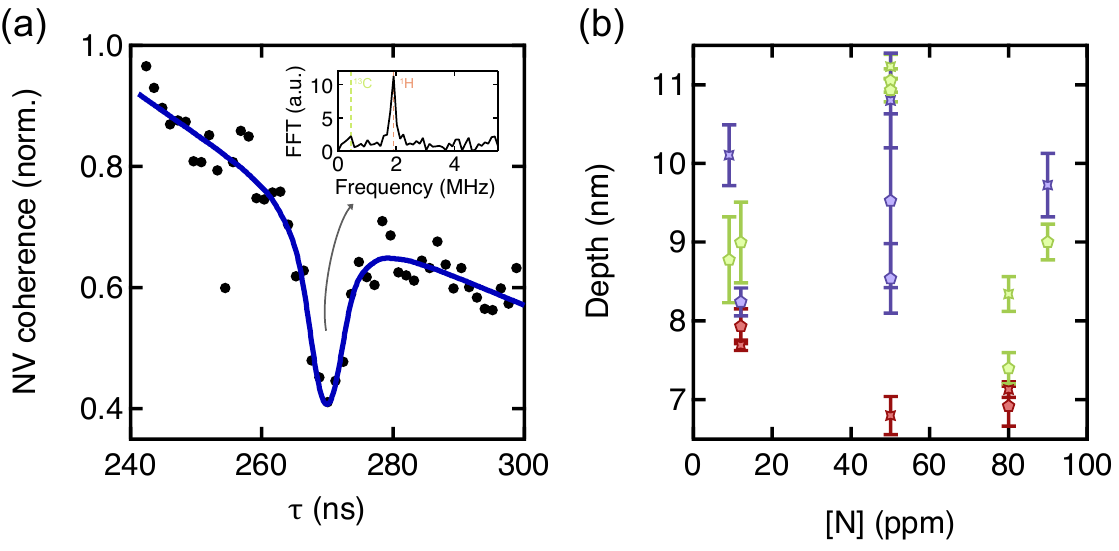}
\caption{\textbf{NV ensemble depth measurements} (a) Example spin decoherence data obtained with an XY16-64 sequence (black dots) and fit (blue line) showing the hydrogen resonance. Inset: FFT of a correlation spectroscopy signal taken on-resonance, showing the hydrogen signal is dominant over the $^{13}$C harmonic. (b) Plot of mean ensemble NV depth $d_{\rm NV}$ versus nitrogen concentration, using the same colour coding as in Fig.~\ref{fig1}. Depths quoted are measured using XY16-64 sequences and the error bars denote either the standard error from the fits or the spread in fit depths given by sequences ranging from 48 to 128 pulses, whichever is larger for a given data point. Note that most but not all samples studied were able to detect a hydrogen signal and those that could not are not included on this plot.}
\label{fig:depth}
\end{figure}

The mean depths of the ensembles created can be measured by taking NV nuclear magnetic resonance (NMR) measurements of a hydrogen target deposited on the diamond surface (in this case viscous immersion oil)~\cite{Pham2016}, see example spectrum showing the appearance of the hydrogen ($^1$H) resonance in Fig.~\ref{fig:depth}(a). For these measurements (and all to follow), we use a widefield microscope optimised for high-sensitivity NV ensemble measurements~\cite{Tetienne2018,Broadway2018a}, except where background fluorescence was problematic (in which case the confocal system was used). A permanent magnet was used to set a magnetic field of 45~mT and was aligned with one set of NV axes.

All samples studied contained natural $^{13}$C abundance (1.1\%), making accurate NV depth determination using XY8 sequences  difficult due to the copresence of a $^{13}$C harmonic with the fundamental $^1$H resonance \cite{Loretz2015}. Where possible, we use the XY16 sequence as it is less sensitive to the problematic fourth $^{13}$C harmonic \cite{Loretz2015}. Even XY16 retains some sensitivity to this harmonic and so all depths quoted should be interpreted as lower bounds of the true mean depth of the ensembles. Correlation spectroscopy~\cite{Staudacher2015} verified that the $^{13}$C harmonic was a relatively minor component of the resonance fit for the shallowest implants (see the FFT in the Fig.~\ref{fig:depth}(a) inset), however was more significant for some of the deeper implants. 

In almost all cases, it was possible to detect a hydrogen signal from immersion oil placed onto the diamond surface using the created layers, however using two of the 16 diamonds implanted could not, indicating that a shallow layer had not been successfully created. This failure could be due to vacancy diffusion into the diamond as increased PL was still observed. It is also possible that, in these diamonds, the yield enhancement from the implantation process was too poor for the hydrogen signal detected by the shallowest NVs to rise above the noise/background given by deeper NVs. Nevertheless, the fact that the majority of samples are able to detect a strong hydrogen signal indicates that the sensing layers are confined close to the surface. From this observation we infer that vacancy diffusion into the diamond under the chosen annealing conditions is not a major factor: substitutional nitrogen is an efficient enough vacancy attractor to dramatically reduce the vacancy diffusion length during annealing, which is important for the success of implantation into type Ib diamond as a method for creating shallow NV layers. Hydrogen signals were detected over the full range of nitrogen densities probed. The results are summarised in the plot Fig.~\ref{fig:depth}(b), with mean ensemble depths ranging from 7 to 11~nm. The depths quoted are given by a 64-pulse sequence in each case, which we take to be a measure of the peak of the NV$^-$ depth distribution~\cite{Healey2021}. The error bars represent the larger of the the uncertainty from the fit and the spread in depth given by measurements with different numbers of pulses (ranging from 48 to 128). Errors due to the copresence of the $^{13}$C harmonic resonance (particularly for deeper implants) and contributions from bulk NV fluorescence (for high-N sectors) are not accounted for, which would cause the underestimation and overestimation of the actual depth respectively. 

The shallowest implants (2.5~keV $^{16}$O and 4~keV $^{31}$P -- represented by the burgundy points), with peak vacancy production predicted below 3~nm from the diamond surface, were measured to have depths between 6.5 and 8~nm, consistent with high-quality N implants of a similar energy~\cite{Pham2016,Healey2021}. This result illustrates two things: firstly that vacancy diffusion into the diamond is much less than order 100~nm observed in the bulk~\cite{Orwa2012,Alsid2019}, which would cause a much deeper mean ensemble depth that would preclude detection of the hydrogen signal. Instead these depths are consistent with the distribution predicted by the SRIM simulation, up to a cut-off introduced by band bending (the same interpretation as for N implants~\cite{Healey2021}). Secondly, however, that these ensembles are (at best) only as shallow as N-implanted ensembles (i.e.\ not shallower) suggests that the high bulk N density does not significantly alter the band bending. 

The 4~keV $^{16}$O and 9~keV $^{31}$P implants (vacancy distribution peaking at 4~nm -- green points) have deeper depth distributions, with most fit depths ranging from 8 to 9~nm. The deepest set of implants, 6~keV $^{16}$O and 11~keV $^{31}$P, (lavender points) had depths measured to be similar to the 4~keV implants, between 8 and 11~nm. As the deeper implants resulted in higher yields on average, these depths may still be in a useful regime and in practice both parameters should be considered alongside one another in determining which implant is appropriate for a particular application. 
\subsection{Ensemble sensitivity}

\begin{figure}
\centering
\includegraphics[width=0.45\textwidth]{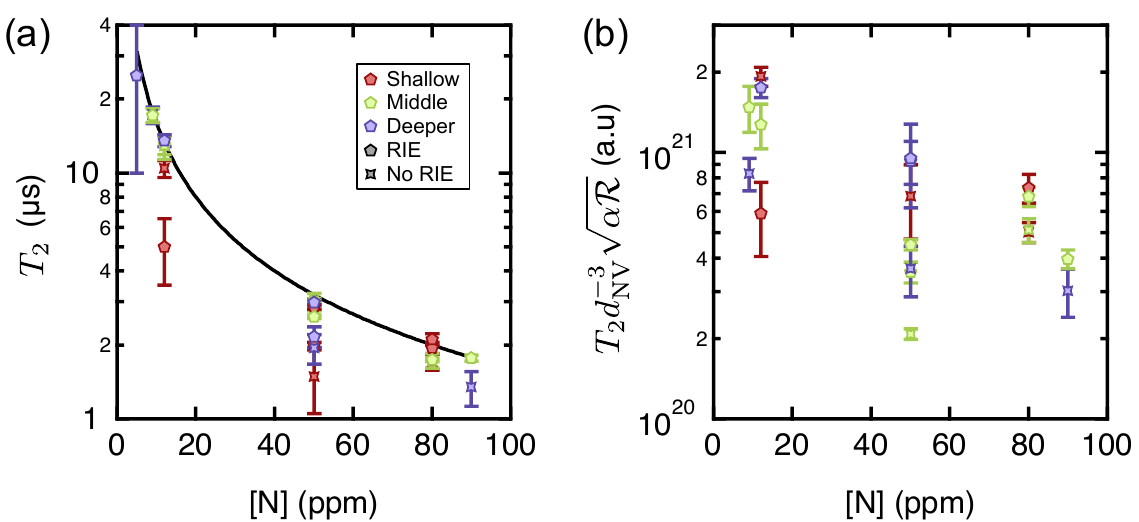}
\caption{\textbf{Assessing NV ensemble quality} (a) Plot of Hahn echo $T_2$ values for the shallow ensembles measured versus the nitrogen content of the sectors. Error bars are the standard errors from the fits to measured decay curves. The solid black line gives the N-limited $T_2$ value given by the equation of Bauch \textit{et al.}~\cite{Bauch2020}. (b) Plot of the figure of merit $T_2 d_{\rm NV}^{-3}\sqrt{\alpha \mathcal R}$ (see text) versus nitrogen content. Error bars are dominantly given by the uncertainty in $T_2$ and $d_{\rm NV}$.}
\label{fig3}
\end{figure}

Although the suitability of a sample to perform a given application will ultimately be heavily dependent on the precise nature of the measurement to take place, we can consider some general figures of merit to gauge the success of the approach. Since most applications of shallow NV ensembles will be concerned with AC signals whose detection can be in principle enhanced through dynamical decoupling, we first measure the Hahn echo $T_2$ of the shallow ensembles, summarised in Fig.~\ref{fig3}(a). We see some evidence for surface-induced decoherence in the shallower implants, with the low-N sector $T_2$ values being longer for deeper implants. At the highest N densities, the various samples are more tightly grouped, implying a $T_2$ close to the N-limited value. The N-limited $T_2$ curve determined by Bauch \textit{et al.}~\cite{Bauch2020} is included as the black line in Fig.~\ref{fig3}(a) to highlight the apparent impact of the surface, however again we stress that the determination of sector [N] may be imperfect and we assume that all sectors in a ``group" have the same N density.

To gauge the overall sensing performance of an NV ensemble (crucially also taking into account the fluorescence of the ensemble, scaling with [NV]), a common figure of merit is the photon shot noise-limited magnetic sensitivity, which for AC fields depends on $T_2$~\cite{Rondin2014,Barry2020,Taylor2008}. As we are concerned here with the detection of rapidly-decaying signals scaling as the cube of the distance between NV and target (e.g. a magnetic noise $B_{\rm RMS}^2 \propto d_{\rm NV}^{-3}$~\cite{Pham2016}) and our ensembles feature different mean depths, we consider instead the minimal figure of merit $T_2 d_{\rm NV}^{-3}\sqrt{\alpha \mathcal{R}}$, which is proportional to the signal-to-noise ratio of a measurement for a given acquisition time. Here $\mathcal{R} \propto$~[NV] is the photon count rate under continuous laser illumination and $\alpha$ is the laser duty cycle for a measurement of the optimal duration $T_2$, both setup-dependent quantities (in this case we use a widefield microscope optimised to measure NV ensembles as a benchmark, as in Ref.~\cite{Healey2020}). We plot this quantity versus [N] in Fig.~\ref{fig3}(b), finding that the spread is partly within error but with an overall tendency for lower-N sectors to perform better. The good performance of low-N sectors, buoyed by their longer $T_2$, is partly a consequence of considering a widefield measurement requiring a long laser pulse duration (5~$\mu$s here) in contrast to confocal microscopy which will have $\alpha \approx 1/T_2$~\cite{Healey2020}, however also reflects the low yields obtained in higher-N sectors and confirms that high bulk N concentration does not appear to aid near-surface NV properties by compensating for electron traps at the surface. The shallowest implants do perform the best on average despite them being most affected by imperfections in the surface preparation which further motivates the pursuit of shallower, stable NV ensembles. We note also that the motivating application of NV-based hyperpolarisation does not rely on shot-noise-limited readout and so a figure of merit scales with [NV] rather than $\sqrt{\rm [NV]}$~\cite{Healey2021} and hence favours the use of more dense ensembles. 

\section{Discussion}
The main limitations of N-implantation for the creation of thick sensing layers is the vacancy-overproduction (e.g.\ peak vacancy production for a 100~keV N implant exceeds the number of implanted ions by a factor of up to 100~\cite{Healey2020}) compared to lower dose implants into (for example) N-rich HPHT diamond and the inability to create layers of arbitrary thickness with a single implantation stage. In the shallow regime, neither of these issues are relevant as the nitrogen depth profiles optimal for the applications discussed are easily attainable with N implantation and the vacancy yield generally is low. Indeed, the localisation of vacancy production to the implanted ions may be beneficial for the purpose of curbing diffusion to the surface by converting vacancies to NV centres most efficiently, although the formation of multi-vacancy clusters may still be problematic~\cite{Tetienne2018}. In view of the above, it would appear that N-implantation is the most suitable technique for creating near-surface NV ensembles. Beginning with a high-quality CVD diamond also carries the benefit of allowing the use of refined doping of the crystal so as to promote NV formation through mechanisms such as vacancy charging as well as Fermi level control~\cite{Luhmann2019}, although these techniques have yet to be convincingly applied in the high-nitrogen, near-surface regime. 

Nevertheless, in this work we have demonstrated that well-confined sensing layers can be produced within 15~nm of the diamond surface via implantation of HPHT diamond. This result shows that vacancy diffusion into the diamond bulk is not the limiting factor for N-to-NV yield near the diamond surface. The low yields may then instead be understood as vacancy diffusion to the nearby surface boundary that acts as a sink. If the surface can be engineered to be vacancy-reflecting, in line with the simulations of R\"acke \textit{et al.}~\cite{Racke2021} N-to-NV yields towards the bulk values of near 10\% may be achievable.

We note that this surface needs to be maintained throughout the annealing process, with maximum temperatures typically ranging from 800-1100$\degree$C~\cite{Tetienne2018}. These temperatures overlap the removal temperatures for common termination species, with oxygen being removed above 600$\degree$C~\cite{Pehrsson2000} and hydrogen above 900$\degree$C~\cite{Riedel2004}. We chose a maximum temperature of 800$\degree$C to mitigate these effects, however even at this temperature and under high vacuum conditions of $\sim 1\times 10^{-6}$~hPa, small amounts of oxygen present in the chamber could disrupt the surface termination. Annealing at higher temperatures comes with the benefit of improving spin properties~\cite{Tetienne2018}, however the surface termination will be even less well controlled and we can expect vacancy diffusion into the diamond to be more significant in this regime, motivating further studies in this area.

The depths and yields measured in this work are broadly similar to typical values measured for shallow ensembles created by N-implantation, indicating that implantation of type Ib diamond could be a cost-effective method of creating shallow NV layers. However, the high bulk nitrogen density present in some sectors does not appear to significantly combat surface-induced band bending meaning the method does not provide any advantages over N-implantation in this regime. N-implantation of electronic-grade diamond is naturally well-suited to creating well-defined shallow NV layers and will not suffer from vacancy diffusion into the diamond, even though our results suggest this is not a major concern regardless. 
\section{Conclusion}
This work has shown that it is possible to create dense, well-confined, shallow NV ensembles via the ion implantation of type Ib HPHT diamond, with yields in the range of those typically achieved using N-implantation. Although we did not find strong evidence for high bulk nitrogen density improving near-surface NV$^-$ charge stability, these results do show that economical production of shallow ensembles is possible using this method. Along with near-surface band-bending, vacancy diffusion to the surface is likely limiting the yield by reducing NV formation efficiency and we speculate that the large spread in measured yields is due to variable surface termination in the diamond samples. A simple oxygen RIE process prior to implantation was not found to dramatically change results by itself, and so focussing on achieving high-quality surface termination during annealing is a logical next step.

The annealing processes conducted here are not expected to be optimal, and the relatively unknown role they have played motivates more systematic studies that could allow improved near-surface ensemble properties. For instance, the use of a higher-temperature anneal has previously been shown to improve the spin properties of shallow ensembles~\cite{Tetienne2018} and the N-to-NV yield could be improved through greater control over the diamond surface termination. Annealing during the implantation process is also an appealing option that has been shown to improve NV yields in the bulk~\cite{Kucsko2018}. Charging vacancies during annealing by introducing shallow electron donors to the diamond crystal may also improve the vacancy yield by limiting the formation of multi-vacancy clusters and perhaps out-diffusion to the surface on electrostatic grounds~\cite{Luhmann2019}. The areas for improvement identified in this work will hopefully allow the creation of shallow ensembles approaching bulk values to be feasible through ion implantation of both electronic grade and type Ib diamond in the future.
\section*{Acknowledgements}
This work was supported by the Australian Research Council (ARC) through Grants CE170100012, DP190101506, and FT200100073. A.J.H. is supported by an Australian Government Research Training Program Scholarship. S.C.S. gratefully acknowledges the support of an Ernst and Grace Matthaei scholarship. This work was performed in part at the Melbourne Centre for Nanofabrication (MCN) in the Victorian Node of the Australian National Fabrication Facility (ANFF).
\bibliographystyle{apsrev} 
\bibliography{shallowhphtbib}

\begin{thebibliography}{42}
\expandafter\ifx\csname natexlab\endcsname\relax\def\natexlab#1{#1}\fi
\expandafter\ifx\csname bibnamefont\endcsname\relax
  \def\bibnamefont#1{#1}\fi
\expandafter\ifx\csname bibfnamefont\endcsname\relax
  \def\bibfnamefont#1{#1}\fi
\expandafter\ifx\csname citenamefont\endcsname\relax
  \def\citenamefont#1{#1}\fi
\expandafter\ifx\csname url\endcsname\relax
  \def\url#1{\texttt{#1}}\fi
\expandafter\ifx\csname urlprefix\endcsname\relax\def\urlprefix{URL }\fi
\providecommand{\bibinfo}[2]{#2}
\providecommand{\eprint}[2][]{\url{#2}}

\bibitem[{\citenamefont{Cole and Hollenberg}(2009)}]{Cole2009}
\bibinfo{author}{\bibfnamefont{J.~H.} \bibnamefont{Cole}} \bibnamefont{and}
  \bibinfo{author}{\bibfnamefont{L.~C.} \bibnamefont{Hollenberg}},
  \bibinfo{title}{{Scanning quantum decoherence microscopy}},
  \bibinfo{journal}{Nanotechnology} \textbf{\bibinfo{volume}{20}},
  \bibinfo{number}{49} (\bibinfo{year}{2009}), \eprint{0811.1913}.

\bibitem[{\citenamefont{Meriles et~al.}(2010)\citenamefont{Meriles, Jiang,
  Goldstein, Hodges, Maze, Lukin, and Capp}}]{Meriles2010}
\bibinfo{author}{\bibfnamefont{C.~A.} \bibnamefont{Meriles}},
  \bibinfo{author}{\bibfnamefont{L.}~\bibnamefont{Jiang}},
  \bibinfo{author}{\bibfnamefont{G.}~\bibnamefont{Goldstein}},
  \bibinfo{author}{\bibfnamefont{J.~S.} \bibnamefont{Hodges}},
  \bibinfo{author}{\bibfnamefont{J.}~\bibnamefont{Maze}},
  \bibinfo{author}{\bibfnamefont{M.~D.} \bibnamefont{Lukin}}, \bibnamefont{and}
  \bibinfo{author}{\bibnamefont{Capp}}, \bibinfo{title}{{Imaging mesoscopic
  nuclear spin noise with a diamond magnetometer}}, \bibinfo{journal}{J. Chem.
  Phys.} \textbf{\bibinfo{volume}{133}}, \bibinfo{number}{124105}
  (\bibinfo{year}{2010}).

\bibitem[{\citenamefont{Staudacher et~al.}(2013)\citenamefont{Staudacher, Shi,
  Pezzagna, Meijer, Du, Meriles, Reinhard, and Wrachtrup}}]{Staudacher2013}
\bibinfo{author}{\bibfnamefont{T.}~\bibnamefont{Staudacher}},
  \bibinfo{author}{\bibfnamefont{F.}~\bibnamefont{Shi}},
  \bibinfo{author}{\bibfnamefont{S.}~\bibnamefont{Pezzagna}},
  \bibinfo{author}{\bibfnamefont{J.}~\bibnamefont{Meijer}},
  \bibinfo{author}{\bibfnamefont{J.}~\bibnamefont{Du}},
  \bibinfo{author}{\bibfnamefont{C.~a.} \bibnamefont{Meriles}},
  \bibinfo{author}{\bibfnamefont{F.}~\bibnamefont{Reinhard}}, \bibnamefont{and}
  \bibinfo{author}{\bibfnamefont{J.}~\bibnamefont{Wrachtrup}},
  \bibinfo{title}{{Nuclear magnetic resonance spectroscopy on a (5-nanometer)
  sample volume.}}, \bibinfo{journal}{Science} \textbf{\bibinfo{volume}{339}},
  \bibinfo{number}{6119} (\bibinfo{year}{2013}).

\bibitem[{\citenamefont{Mamin et~al.}(2013)\citenamefont{Mamin, Kim, Sherwood,
  Rettner, Ohno, Awschalom, and Rugar}}]{Mamin2013}
\bibinfo{author}{\bibfnamefont{H.~J.} \bibnamefont{Mamin}},
  \bibinfo{author}{\bibfnamefont{M.}~\bibnamefont{Kim}},
  \bibinfo{author}{\bibfnamefont{M.~H.} \bibnamefont{Sherwood}},
  \bibinfo{author}{\bibfnamefont{C.~T.} \bibnamefont{Rettner}},
  \bibinfo{author}{\bibfnamefont{K.}~\bibnamefont{Ohno}},
  \bibinfo{author}{\bibfnamefont{D.~D.} \bibnamefont{Awschalom}},
  \bibnamefont{and} \bibinfo{author}{\bibfnamefont{D.}~\bibnamefont{Rugar}},
  \bibinfo{title}{{Nanoscale Nuclear Magnetic Resonance with a Nitrogen-Vacancy
  Spin Sensor}}, \bibinfo{journal}{Science} \textbf{\bibinfo{volume}{339}},
  \bibinfo{number}{February} (\bibinfo{year}{2013}).

\bibitem[{\citenamefont{Loretz et~al.}(2014)\citenamefont{Loretz, Pezzagna,
  Meijer, and Degen}}]{Loretz2014}
\bibinfo{author}{\bibfnamefont{M.}~\bibnamefont{Loretz}},
  \bibinfo{author}{\bibfnamefont{S.}~\bibnamefont{Pezzagna}},
  \bibinfo{author}{\bibfnamefont{J.}~\bibnamefont{Meijer}}, \bibnamefont{and}
  \bibinfo{author}{\bibfnamefont{C.~L.} \bibnamefont{Degen}},
  \bibinfo{title}{{Nanoscale nuclear magnetic resonance with a 1.9-nm-deep
  nitrogen-vacancy sensor}}, \bibinfo{journal}{Appl. Phys. Lett.}
  \textbf{\bibinfo{volume}{104}}, \bibinfo{number}{3} (\bibinfo{year}{2014}).

\bibitem[{\citenamefont{Muller et~al.}(2014)\citenamefont{Muller, Kong, Cai,
  Melentijevic, Stacey, Markham, Twitchen, Isoya, Pezzagna, Meijer
  et~al.}}]{Muller2014}
\bibinfo{author}{\bibfnamefont{C.}~\bibnamefont{Muller}},
  \bibinfo{author}{\bibfnamefont{X.}~\bibnamefont{Kong}},
  \bibinfo{author}{\bibfnamefont{J.-M.} \bibnamefont{Cai}},
  \bibinfo{author}{\bibfnamefont{K.}~\bibnamefont{Melentijevic}},
  \bibinfo{author}{\bibfnamefont{A.}~\bibnamefont{Stacey}},
  \bibinfo{author}{\bibfnamefont{M.}~\bibnamefont{Markham}},
  \bibinfo{author}{\bibfnamefont{D.}~\bibnamefont{Twitchen}},
  \bibinfo{author}{\bibfnamefont{J.}~\bibnamefont{Isoya}},
  \bibinfo{author}{\bibfnamefont{S.}~\bibnamefont{Pezzagna}},
  \bibinfo{author}{\bibfnamefont{J.}~\bibnamefont{Meijer}},
  \bibnamefont{et~al.}, \bibinfo{title}{{Nuclear magnetic resonance
  spectroscopy with single spin sensitivity}}, \bibinfo{journal}{Nat. Commun.}
  \textbf{\bibinfo{volume}{5}}, \bibinfo{number}{4703} (\bibinfo{year}{2014}).

\bibitem[{\citenamefont{Devience et~al.}(2015)\citenamefont{Devience, Pham,
  Lovchinsky, Sushkov, Bar-Gill, Belthangady, Casola, Corbett, Zhang, Lukin
  et~al.}}]{Devience2015}
\bibinfo{author}{\bibfnamefont{S.~J.} \bibnamefont{Devience}},
  \bibinfo{author}{\bibfnamefont{L.~M.} \bibnamefont{Pham}},
  \bibinfo{author}{\bibfnamefont{I.}~\bibnamefont{Lovchinsky}},
  \bibinfo{author}{\bibfnamefont{A.~O.} \bibnamefont{Sushkov}},
  \bibinfo{author}{\bibfnamefont{N.}~\bibnamefont{Bar-Gill}},
  \bibinfo{author}{\bibfnamefont{C.}~\bibnamefont{Belthangady}},
  \bibinfo{author}{\bibfnamefont{F.}~\bibnamefont{Casola}},
  \bibinfo{author}{\bibfnamefont{M.}~\bibnamefont{Corbett}},
  \bibinfo{author}{\bibfnamefont{H.}~\bibnamefont{Zhang}},
  \bibinfo{author}{\bibfnamefont{M.}~\bibnamefont{Lukin}},
  \bibnamefont{et~al.}, \bibinfo{title}{{Nanoscale NMR spectroscopy and imaging
  of multiple nuclear species}}, \bibinfo{journal}{Nat. Nanotechnol.}
  \textbf{\bibinfo{volume}{10}} (\bibinfo{year}{2015}).

\bibitem[{\citenamefont{Lovchinsky et~al.}(2016)\citenamefont{Lovchinsky,
  Sushkov, Urbach, de~Leon, Choi, {De Greve}, Evans, Gertner, Bersin, Muller
  et~al.}}]{Lovchinsky2016}
\bibinfo{author}{\bibfnamefont{I.}~\bibnamefont{Lovchinsky}},
  \bibinfo{author}{\bibfnamefont{A.~O.} \bibnamefont{Sushkov}},
  \bibinfo{author}{\bibfnamefont{E.}~\bibnamefont{Urbach}},
  \bibinfo{author}{\bibfnamefont{N.~P.} \bibnamefont{de~Leon}},
  \bibinfo{author}{\bibfnamefont{S.}~\bibnamefont{Choi}},
  \bibinfo{author}{\bibfnamefont{K.}~\bibnamefont{{De Greve}}},
  \bibinfo{author}{\bibfnamefont{R.}~\bibnamefont{Evans}},
  \bibinfo{author}{\bibfnamefont{R.}~\bibnamefont{Gertner}},
  \bibinfo{author}{\bibfnamefont{E.}~\bibnamefont{Bersin}},
  \bibinfo{author}{\bibfnamefont{C.}~\bibnamefont{Muller}},
  \bibnamefont{et~al.}, \bibinfo{title}{{Nuclear magnetic resonance detection
  and spectroscopy of single proteins using quantum logic}},
  \bibinfo{journal}{Science} \textbf{\bibinfo{volume}{351}},
  \bibinfo{number}{6275} (\bibinfo{year}{2016}).

\bibitem[{\citenamefont{Shi et~al.}(2015)\citenamefont{Shi, Zhang, Sun, Wang,
  Rong, Chen, Ju, Reinhard, Chen, Wrachtrup et~al.}}]{Shi2015}
\bibinfo{author}{\bibfnamefont{F.}~\bibnamefont{Shi}},
  \bibinfo{author}{\bibfnamefont{Q.}~\bibnamefont{Zhang}},
  \bibinfo{author}{\bibfnamefont{H.}~\bibnamefont{Sun}},
  \bibinfo{author}{\bibfnamefont{J.}~\bibnamefont{Wang}},
  \bibinfo{author}{\bibfnamefont{X.}~\bibnamefont{Rong}},
  \bibinfo{author}{\bibfnamefont{M.}~\bibnamefont{Chen}},
  \bibinfo{author}{\bibfnamefont{C.}~\bibnamefont{Ju}},
  \bibinfo{author}{\bibfnamefont{F.}~\bibnamefont{Reinhard}},
  \bibinfo{author}{\bibfnamefont{H.}~\bibnamefont{Chen}},
  \bibinfo{author}{\bibfnamefont{J.}~\bibnamefont{Wrachtrup}},
  \bibnamefont{et~al.}, \bibinfo{title}{{Single-protein spin resonance
  spectroscopy under ambient conditions}}, \bibinfo{journal}{Science}
  \textbf{\bibinfo{volume}{347}}, \bibinfo{number}{6226}
  (\bibinfo{year}{2015}).

\bibitem[{\citenamefont{Shagieva et~al.}(2018)\citenamefont{Shagieva, Zaiser,
  Dasari, Stohr, Denisenko, Reuter, Meriles, and Wrachtrup}}]{Shagieva2018}
\bibinfo{author}{\bibfnamefont{F.}~\bibnamefont{Shagieva}},
  \bibinfo{author}{\bibfnamefont{S.}~\bibnamefont{Zaiser}},
  \bibinfo{author}{\bibfnamefont{D.~B.~R.} \bibnamefont{Dasari}},
  \bibinfo{author}{\bibfnamefont{R.}~\bibnamefont{Stohr}},
  \bibinfo{author}{\bibfnamefont{A.}~\bibnamefont{Denisenko}},
  \bibinfo{author}{\bibfnamefont{R.}~\bibnamefont{Reuter}},
  \bibinfo{author}{\bibfnamefont{C.~A.} \bibnamefont{Meriles}},
  \bibnamefont{and}
  \bibinfo{author}{\bibfnamefont{J.}~\bibnamefont{Wrachtrup}},
  \bibinfo{title}{{Microwave-Assisted Cross-Polarization of Nuclear Spin
  Ensembles from Optically Pumped Nitrogen-Vacancy Centers in Diamond}},
  \bibinfo{journal}{Nano Lett.} \textbf{\bibinfo{volume}{18}}
  (\bibinfo{year}{2018}).

\bibitem[{\citenamefont{Fernandez-Acebal
  et~al.}(2018)\citenamefont{Fernandez-Acebal, Rosolio, Scheuer, Muller,
  Muller, Schmitt, Mcguinness, Schwarz, Chen, Retzker
  et~al.}}]{Fernandez-Acebal2018}
\bibinfo{author}{\bibfnamefont{P.}~\bibnamefont{Fernandez-Acebal}},
  \bibinfo{author}{\bibfnamefont{O.}~\bibnamefont{Rosolio}},
  \bibinfo{author}{\bibfnamefont{J.}~\bibnamefont{Scheuer}},
  \bibinfo{author}{\bibfnamefont{C.}~\bibnamefont{Muller}},
  \bibinfo{author}{\bibfnamefont{S.}~\bibnamefont{Muller}},
  \bibinfo{author}{\bibfnamefont{S.}~\bibnamefont{Schmitt}},
  \bibinfo{author}{\bibfnamefont{L.~P.} \bibnamefont{Mcguinness}},
  \bibinfo{author}{\bibfnamefont{I.}~\bibnamefont{Schwarz}},
  \bibinfo{author}{\bibfnamefont{W.}~\bibnamefont{Chen}},
  \bibinfo{author}{\bibfnamefont{A.}~\bibnamefont{Retzker}},
  \bibnamefont{et~al.}, \bibinfo{title}{{Toward Hyperpolarization of Oil
  Molecules via Single Nitrogen Vacancy Centers in Diamond}},
  \bibinfo{journal}{Nano Lett.} \textbf{\bibinfo{volume}{18}}
  (\bibinfo{year}{2018}).

\bibitem[{\citenamefont{Broadway
  et~al.}(2018{\natexlab{a}})\citenamefont{Broadway, Tetienne, Stacey, Wood,
  Simpson, Hall, and Hollenberg}}]{Broadway2018a}
\bibinfo{author}{\bibfnamefont{D.~A.} \bibnamefont{Broadway}},
  \bibinfo{author}{\bibfnamefont{J.-p.} \bibnamefont{Tetienne}},
  \bibinfo{author}{\bibfnamefont{A.}~\bibnamefont{Stacey}},
  \bibinfo{author}{\bibfnamefont{J.~D.~A.} \bibnamefont{Wood}},
  \bibinfo{author}{\bibfnamefont{D.~A.} \bibnamefont{Simpson}},
  \bibinfo{author}{\bibfnamefont{L.~T.} \bibnamefont{Hall}}, \bibnamefont{and}
  \bibinfo{author}{\bibfnamefont{L.~C.~L.} \bibnamefont{Hollenberg}},
  \bibinfo{title}{{Quantum probe hyperpolarisation of molecular nuclear
  spins}}, \bibinfo{journal}{Nat. Commun.} \textbf{\bibinfo{volume}{9}},
  \bibinfo{number}{1246} (\bibinfo{year}{2018}{\natexlab{a}}).

\bibitem[{\citenamefont{Tetienne et~al.}(2021)\citenamefont{Tetienne, Hall,
  Healey, White, Sani, Separovic, and Hollenberg}}]{Tetienne2021}
\bibinfo{author}{\bibfnamefont{J.-P.} \bibnamefont{Tetienne}},
  \bibinfo{author}{\bibfnamefont{L.~T.} \bibnamefont{Hall}},
  \bibinfo{author}{\bibfnamefont{A.~J.} \bibnamefont{Healey}},
  \bibinfo{author}{\bibfnamefont{G.~A.~L.} \bibnamefont{White}},
  \bibinfo{author}{\bibfnamefont{M.-A.} \bibnamefont{Sani}},
  \bibinfo{author}{\bibfnamefont{F.}~\bibnamefont{Separovic}},
  \bibnamefont{and} \bibinfo{author}{\bibfnamefont{L.~C.~L.}
  \bibnamefont{Hollenberg}}, \bibinfo{title}{{Prospects for nuclear spin
  hyperpolarization of molecular samples using nitrogen-vacancy centers in
  diamond}}, \bibinfo{journal}{Phys. Rev. B} \textbf{\bibinfo{volume}{103}},
  \bibinfo{number}{014434} (\bibinfo{year}{2021}).

\bibitem[{\citenamefont{Collins}(2002)}]{Collins2002}
\bibinfo{author}{\bibfnamefont{A.~T.} \bibnamefont{Collins}},
  \bibinfo{title}{{The Fermi level in diamond}}, \bibinfo{journal}{J. Phys.
  Condens. Matter} \textbf{\bibinfo{volume}{14}}, \bibinfo{number}{3743}
  (\bibinfo{year}{2002}).

\bibitem[{\citenamefont{Healey et~al.}(2021)\citenamefont{Healey, Hall, White,
  Teraji, Sani, Separovic, Tetienne, and Hollenberg}}]{Healey2021}
\bibinfo{author}{\bibfnamefont{A.~J.} \bibnamefont{Healey}},
  \bibinfo{author}{\bibfnamefont{L.~T.} \bibnamefont{Hall}},
  \bibinfo{author}{\bibfnamefont{G.~A.~L.} \bibnamefont{White}},
  \bibinfo{author}{\bibfnamefont{T.}~\bibnamefont{Teraji}},
  \bibinfo{author}{\bibfnamefont{M.}~\bibnamefont{Sani}},
  \bibinfo{author}{\bibfnamefont{F.}~\bibnamefont{Separovic}},
  \bibinfo{author}{\bibfnamefont{J.-P.} \bibnamefont{Tetienne}},
  \bibnamefont{and} \bibinfo{author}{\bibfnamefont{L.~C.~L.}
  \bibnamefont{Hollenberg}}, \bibinfo{title}{{Polarization Transfer to External
  Nuclear Spins Using Ensembles of Nitrogen-Vacancy Centers}},
  \bibinfo{journal}{Phys. Rev. Appl.} \textbf{\bibinfo{volume}{15}},
  \bibinfo{number}{054052} (\bibinfo{year}{2021}).

\bibitem[{\citenamefont{Rizzato et~al.}(2022)\citenamefont{Rizzato, Bruckmaier,
  Liu, Glaser, and Bucher}}]{Rizzato2022}
\bibinfo{author}{\bibfnamefont{R.}~\bibnamefont{Rizzato}},
  \bibinfo{author}{\bibfnamefont{F.}~\bibnamefont{Bruckmaier}},
  \bibinfo{author}{\bibfnamefont{K.~S.} \bibnamefont{Liu}},
  \bibinfo{author}{\bibfnamefont{S.~J.} \bibnamefont{Glaser}},
  \bibnamefont{and} \bibinfo{author}{\bibfnamefont{D.~B.}
  \bibnamefont{Bucher}}, \bibinfo{title}{{Polarization Transfer from Optically
  Pumped Ensembles of N- V Centers to Multinuclear Spin Baths}},
  \bibinfo{journal}{Phys. Rev. Appl.} \textbf{\bibinfo{volume}{17}},
  \bibinfo{number}{024067} (\bibinfo{year}{2022}).

\bibitem[{\citenamefont{Steinert et~al.}(2013)\citenamefont{Steinert, Ziem,
  Hall, Zappe, Schweikert, Gotz, Aird, Balasubramanian, Hollenberg, and
  Wrachtrup}}]{Steinert2013}
\bibinfo{author}{\bibfnamefont{S.}~\bibnamefont{Steinert}},
  \bibinfo{author}{\bibfnamefont{F.}~\bibnamefont{Ziem}},
  \bibinfo{author}{\bibfnamefont{L.~T.} \bibnamefont{Hall}},
  \bibinfo{author}{\bibfnamefont{A.}~\bibnamefont{Zappe}},
  \bibinfo{author}{\bibfnamefont{M.}~\bibnamefont{Schweikert}},
  \bibinfo{author}{\bibfnamefont{N.}~\bibnamefont{Gotz}},
  \bibinfo{author}{\bibfnamefont{A.}~\bibnamefont{Aird}},
  \bibinfo{author}{\bibfnamefont{G.}~\bibnamefont{Balasubramanian}},
  \bibinfo{author}{\bibfnamefont{L.}~\bibnamefont{Hollenberg}},
  \bibnamefont{and}
  \bibinfo{author}{\bibfnamefont{J.}~\bibnamefont{Wrachtrup}},
  \bibinfo{title}{{Magnetic spin imaging under ambient conditions with
  sub-cellular resolution}}, \bibinfo{journal}{Nat. Commun.}
  \textbf{\bibinfo{volume}{4}}, \bibinfo{number}{1607} (\bibinfo{year}{2013}).

\bibitem[{\citenamefont{Simpson et~al.}(2017)\citenamefont{Simpson, Ryan, Hall,
  Panchenko, Drew, Petrou, Donnelly, Mulvaney, and Hollenberg}}]{Simpson2017}
\bibinfo{author}{\bibfnamefont{D.~A.} \bibnamefont{Simpson}},
  \bibinfo{author}{\bibfnamefont{R.~G.} \bibnamefont{Ryan}},
  \bibinfo{author}{\bibfnamefont{L.~T.} \bibnamefont{Hall}},
  \bibinfo{author}{\bibfnamefont{E.}~\bibnamefont{Panchenko}},
  \bibinfo{author}{\bibfnamefont{S.~C.} \bibnamefont{Drew}},
  \bibinfo{author}{\bibfnamefont{S.}~\bibnamefont{Petrou}},
  \bibinfo{author}{\bibfnamefont{P.~S.} \bibnamefont{Donnelly}},
  \bibinfo{author}{\bibfnamefont{P.}~\bibnamefont{Mulvaney}}, \bibnamefont{and}
  \bibinfo{author}{\bibfnamefont{L.~C.~L.} \bibnamefont{Hollenberg}},
  \bibinfo{title}{{Electron paramagnetic resonance microscopy using spins in
  diamond under ambient conditions}}, \bibinfo{journal}{Nat. Commun.}
  \textbf{\bibinfo{volume}{8}}, \bibinfo{number}{458} (\bibinfo{year}{2017}).

\bibitem[{\citenamefont{McCoey et~al.}(2020)\citenamefont{McCoey, Matsuoka,
  de~Gille, Hall, Shaw, Tetienne, Kisailus, Hollenberg, and
  Simpson}}]{McCoey2020}
\bibinfo{author}{\bibfnamefont{J.~M.} \bibnamefont{McCoey}},
  \bibinfo{author}{\bibfnamefont{M.}~\bibnamefont{Matsuoka}},
  \bibinfo{author}{\bibfnamefont{R.~W.} \bibnamefont{de~Gille}},
  \bibinfo{author}{\bibfnamefont{L.~T.} \bibnamefont{Hall}},
  \bibinfo{author}{\bibfnamefont{J.~A.} \bibnamefont{Shaw}},
  \bibinfo{author}{\bibfnamefont{J.-p.} \bibnamefont{Tetienne}},
  \bibinfo{author}{\bibfnamefont{D.}~\bibnamefont{Kisailus}},
  \bibinfo{author}{\bibfnamefont{L.~C.~L.} \bibnamefont{Hollenberg}},
  \bibnamefont{and} \bibinfo{author}{\bibfnamefont{D.~A.}
  \bibnamefont{Simpson}}, \bibinfo{title}{{Quantum Magnetic Imaging of Iron
  Biomineralization in Teeth of the Chiton Acanthopleura hirtosa}},
  \bibinfo{journal}{Small Methods} \textbf{\bibinfo{volume}{1900754}}
  (\bibinfo{year}{2020}).

\bibitem[{\citenamefont{Ziem et~al.}(2019)\citenamefont{Ziem, Garsi, Fedder,
  and Wrachtrup}}]{Ziem2019}
\bibinfo{author}{\bibfnamefont{F.}~\bibnamefont{Ziem}},
  \bibinfo{author}{\bibfnamefont{M.}~\bibnamefont{Garsi}},
  \bibinfo{author}{\bibfnamefont{H.}~\bibnamefont{Fedder}}, \bibnamefont{and}
  \bibinfo{author}{\bibfnamefont{J.}~\bibnamefont{Wrachtrup}},
  \bibinfo{title}{{Quantitative nanoscale MRI with a wide field of view}},
  \bibinfo{journal}{Sci. Rep.} \textbf{\bibinfo{volume}{9}},
  \bibinfo{number}{12166} (\bibinfo{year}{2019}).

\bibitem[{\citenamefont{Bertelli et~al.}(2020)\citenamefont{Bertelli,
  Carmiggelt, Yu, Simon, Pothoven, Bauer, Blanter, Aarts, and
  Sar}}]{Bertelli2020}
\bibinfo{author}{\bibfnamefont{I.}~\bibnamefont{Bertelli}},
  \bibinfo{author}{\bibfnamefont{J.~J.} \bibnamefont{Carmiggelt}},
  \bibinfo{author}{\bibfnamefont{T.}~\bibnamefont{Yu}},
  \bibinfo{author}{\bibfnamefont{B.~G.} \bibnamefont{Simon}},
  \bibinfo{author}{\bibfnamefont{C.~C.} \bibnamefont{Pothoven}},
  \bibinfo{author}{\bibfnamefont{G.~E.~W.} \bibnamefont{Bauer}},
  \bibinfo{author}{\bibfnamefont{Y.~M.} \bibnamefont{Blanter}},
  \bibinfo{author}{\bibfnamefont{J.}~\bibnamefont{Aarts}}, \bibnamefont{and}
  \bibinfo{author}{\bibfnamefont{T.~V.~D.} \bibnamefont{Sar}},
  \bibinfo{title}{{Magnetic resonance imaging of spin-wave transport and
  interference in a magnetic insulator}}, \bibinfo{journal}{Sci. Adv.}
  \textbf{\bibinfo{volume}{6}}, \bibinfo{number}{eabd3556}
  (\bibinfo{year}{2020}).

\bibitem[{\citenamefont{Broadway
  et~al.}(2018{\natexlab{b}})\citenamefont{Broadway, Dontschuk, Tsai, Lillie,
  Lew, Mccallum, Johnson, Doherty, Stacey, Hollenberg et~al.}}]{Broadway2018b}
\bibinfo{author}{\bibfnamefont{D.~A.} \bibnamefont{Broadway}},
  \bibinfo{author}{\bibfnamefont{N.}~\bibnamefont{Dontschuk}},
  \bibinfo{author}{\bibfnamefont{A.}~\bibnamefont{Tsai}},
  \bibinfo{author}{\bibfnamefont{S.~E.} \bibnamefont{Lillie}},
  \bibinfo{author}{\bibfnamefont{C.~T.} \bibnamefont{Lew}},
  \bibinfo{author}{\bibfnamefont{J.~C.} \bibnamefont{Mccallum}},
  \bibinfo{author}{\bibfnamefont{B.~C.} \bibnamefont{Johnson}},
  \bibinfo{author}{\bibfnamefont{M.~W.} \bibnamefont{Doherty}},
  \bibinfo{author}{\bibfnamefont{A.}~\bibnamefont{Stacey}},
  \bibinfo{author}{\bibfnamefont{L.~C.~L.} \bibnamefont{Hollenberg}},
  \bibnamefont{et~al.}, \bibinfo{title}{{Spatial mapping of band bending in
  semiconductor devices using in situ quantum sensors}}, \bibinfo{journal}{Nat.
  Electron.} \textbf{\bibinfo{volume}{1}} (\bibinfo{year}{2018}{\natexlab{b}}).

\bibitem[{\citenamefont{Pezzagna et~al.}(2010)\citenamefont{Pezzagna, Naydenov,
  Jelezko, Wrachtrup, and Meijer}}]{Pezzagna2010}
\bibinfo{author}{\bibfnamefont{S.}~\bibnamefont{Pezzagna}},
  \bibinfo{author}{\bibfnamefont{B.}~\bibnamefont{Naydenov}},
  \bibinfo{author}{\bibfnamefont{F.}~\bibnamefont{Jelezko}},
  \bibinfo{author}{\bibfnamefont{J.}~\bibnamefont{Wrachtrup}},
  \bibnamefont{and} \bibinfo{author}{\bibfnamefont{J.}~\bibnamefont{Meijer}},
  \bibinfo{title}{{Creation efficiency of nitrogen-vacancy centres in
  diamond}}, \bibinfo{journal}{New J. Phys.} \textbf{\bibinfo{volume}{12}},
  \bibinfo{number}{065017} (\bibinfo{year}{2010}).

\bibitem[{\citenamefont{Racke et~al.}(2021)\citenamefont{Racke, Pietzonka,
  Meijer, Spemann, and Wunderlich}}]{Racke2021}
\bibinfo{author}{\bibfnamefont{P.}~\bibnamefont{Racke}},
  \bibinfo{author}{\bibfnamefont{L.}~\bibnamefont{Pietzonka}},
  \bibinfo{author}{\bibfnamefont{J.}~\bibnamefont{Meijer}},
  \bibinfo{author}{\bibfnamefont{D.}~\bibnamefont{Spemann}}, \bibnamefont{and}
  \bibinfo{author}{\bibfnamefont{R.}~\bibnamefont{Wunderlich}},
  \bibinfo{title}{{Vacancy diffusion and nitrogen-vacancy center formation near
  the diamond surface}}, \bibinfo{journal}{Appl. Phys. Lett.}
  \textbf{\bibinfo{volume}{118}}, \bibinfo{number}{204003}
  (\bibinfo{year}{2021}).

\bibitem[{\citenamefont{Huang et~al.}(2013)\citenamefont{Huang, Li, Santori,
  Acosta, Faraon, Ishikawa, Wu, Winston, Williams, and Beausoleil}}]{Huang2013}
\bibinfo{author}{\bibfnamefont{Z.}~\bibnamefont{Huang}},
  \bibinfo{author}{\bibfnamefont{W.~D.} \bibnamefont{Li}},
  \bibinfo{author}{\bibfnamefont{C.}~\bibnamefont{Santori}},
  \bibinfo{author}{\bibfnamefont{V.~M.} \bibnamefont{Acosta}},
  \bibinfo{author}{\bibfnamefont{A.}~\bibnamefont{Faraon}},
  \bibinfo{author}{\bibfnamefont{T.}~\bibnamefont{Ishikawa}},
  \bibinfo{author}{\bibfnamefont{W.}~\bibnamefont{Wu}},
  \bibinfo{author}{\bibfnamefont{D.}~\bibnamefont{Winston}},
  \bibinfo{author}{\bibfnamefont{R.~S.} \bibnamefont{Williams}},
  \bibnamefont{and} \bibinfo{author}{\bibfnamefont{R.~G.}
  \bibnamefont{Beausoleil}}, \bibinfo{title}{{Diamond nitrogen-vacancy centers
  created by scanning focused helium ion beam and annealing}},
  \bibinfo{journal}{Applied Physics Letters} \textbf{\bibinfo{volume}{103}},
  \bibinfo{number}{8} (\bibinfo{year}{2013}).

\bibitem[{\citenamefont{Mccloskey et~al.}(2014)\citenamefont{Mccloskey, Fox,
  Hara, Usov, Scanlan, Mcevoy, Duesberg, Cross, Donegan, Mccloskey
  et~al.}}]{Mccloskey2014}
\bibinfo{author}{\bibfnamefont{D.}~\bibnamefont{Mccloskey}},
  \bibinfo{author}{\bibfnamefont{D.}~\bibnamefont{Fox}},
  \bibinfo{author}{\bibfnamefont{N.~O.} \bibnamefont{Hara}},
  \bibinfo{author}{\bibfnamefont{V.}~\bibnamefont{Usov}},
  \bibinfo{author}{\bibfnamefont{D.}~\bibnamefont{Scanlan}},
  \bibinfo{author}{\bibfnamefont{N.}~\bibnamefont{Mcevoy}},
  \bibinfo{author}{\bibfnamefont{G.~S.} \bibnamefont{Duesberg}},
  \bibinfo{author}{\bibfnamefont{G.~L.~W.} \bibnamefont{Cross}},
  \bibinfo{author}{\bibfnamefont{J.~F.} \bibnamefont{Donegan}},
  \bibinfo{author}{\bibfnamefont{D.}~\bibnamefont{Mccloskey}},
  \bibnamefont{et~al.}, \bibinfo{title}{{Helium ion microscope generated
  nitrogen-vacancy centres in type Ib diamond}}, \bibinfo{journal}{Appl. Phys.
  Lett.} \textbf{\bibinfo{volume}{104}}, \bibinfo{number}{031109}
  (\bibinfo{year}{2014}).

\bibitem[{\citenamefont{Fescenko et~al.}(2019)\citenamefont{Fescenko, Laraoui,
  Smits, Mosavian, Kehayias, Seto, Bougas, Jarmola, and Acosta}}]{Fescenko2019}
\bibinfo{author}{\bibfnamefont{I.}~\bibnamefont{Fescenko}},
  \bibinfo{author}{\bibfnamefont{A.}~\bibnamefont{Laraoui}},
  \bibinfo{author}{\bibfnamefont{J.}~\bibnamefont{Smits}},
  \bibinfo{author}{\bibfnamefont{N.}~\bibnamefont{Mosavian}},
  \bibinfo{author}{\bibfnamefont{P.}~\bibnamefont{Kehayias}},
  \bibinfo{author}{\bibfnamefont{J.}~\bibnamefont{Seto}},
  \bibinfo{author}{\bibfnamefont{L.}~\bibnamefont{Bougas}},
  \bibinfo{author}{\bibfnamefont{A.}~\bibnamefont{Jarmola}}, \bibnamefont{and}
  \bibinfo{author}{\bibfnamefont{V.~M.} \bibnamefont{Acosta}},
  \bibinfo{title}{{Diamond Magnetic Microscopy of Malarial Hemozoin
  Nanocrystals}}, \bibinfo{journal}{Phys. Rev. Appl.}
  \textbf{\bibinfo{volume}{11}} (\bibinfo{year}{2019}).

\bibitem[{\citenamefont{Healey et~al.}(2020)\citenamefont{Healey, Stacey,
  Johnson, Broadway, Teraji, Simpson, Tetienne, and Hollenberg}}]{Healey2020}
\bibinfo{author}{\bibfnamefont{A.~J.} \bibnamefont{Healey}},
  \bibinfo{author}{\bibfnamefont{A.}~\bibnamefont{Stacey}},
  \bibinfo{author}{\bibfnamefont{B.~C.} \bibnamefont{Johnson}},
  \bibinfo{author}{\bibfnamefont{D.~A.} \bibnamefont{Broadway}},
  \bibinfo{author}{\bibfnamefont{T.}~\bibnamefont{Teraji}},
  \bibinfo{author}{\bibfnamefont{D.~A.} \bibnamefont{Simpson}},
  \bibinfo{author}{\bibfnamefont{J.}~\bibnamefont{Tetienne}}, \bibnamefont{and}
  \bibinfo{author}{\bibfnamefont{L.~C.~L.} \bibnamefont{Hollenberg}},
  \bibinfo{title}{{Comparison of different methods of nitrogen-vacancy layer
  formation in diamond for wide-field quantum microscopy}},
  \bibinfo{journal}{Phys. Rev. Mater.} \textbf{\bibinfo{volume}{4}},
  \bibinfo{number}{104605} (\bibinfo{year}{2020}).

\bibitem[{\citenamefont{Orwa et~al.}(2012)\citenamefont{Orwa, Ganesan, Newnham,
  Santori, Barclay, Fu, Beausoleil, Aharonovich, Fairchild, Olivero
  et~al.}}]{Orwa2012}
\bibinfo{author}{\bibfnamefont{J.~O.} \bibnamefont{Orwa}},
  \bibinfo{author}{\bibfnamefont{K.}~\bibnamefont{Ganesan}},
  \bibinfo{author}{\bibfnamefont{J.}~\bibnamefont{Newnham}},
  \bibinfo{author}{\bibfnamefont{C.}~\bibnamefont{Santori}},
  \bibinfo{author}{\bibfnamefont{P.}~\bibnamefont{Barclay}},
  \bibinfo{author}{\bibfnamefont{K.~M.~C.} \bibnamefont{Fu}},
  \bibinfo{author}{\bibfnamefont{R.~G.} \bibnamefont{Beausoleil}},
  \bibinfo{author}{\bibfnamefont{I.}~\bibnamefont{Aharonovich}},
  \bibinfo{author}{\bibfnamefont{B.~A.} \bibnamefont{Fairchild}},
  \bibinfo{author}{\bibfnamefont{P.}~\bibnamefont{Olivero}},
  \bibnamefont{et~al.}, \bibinfo{title}{{An upper limit on the lateral vacancy
  diffusion length in diamond}}, \bibinfo{journal}{Diam. Relat. Mater.}
  \textbf{\bibinfo{volume}{24}}, \bibinfo{number}{6-10} (\bibinfo{year}{2012}).

\bibitem[{\citenamefont{Alsid et~al.}(2019)\citenamefont{Alsid, Barry, Pham,
  Schloss, Keeffe, Cappellaro, and Braje}}]{Alsid2019}
\bibinfo{author}{\bibfnamefont{S.~T.} \bibnamefont{Alsid}},
  \bibinfo{author}{\bibfnamefont{J.~F.} \bibnamefont{Barry}},
  \bibinfo{author}{\bibfnamefont{L.~M.} \bibnamefont{Pham}},
  \bibinfo{author}{\bibfnamefont{J.~M.} \bibnamefont{Schloss}},
  \bibinfo{author}{\bibfnamefont{M.~F.~O.} \bibnamefont{Keeffe}},
  \bibinfo{author}{\bibfnamefont{P.}~\bibnamefont{Cappellaro}},
  \bibnamefont{and} \bibinfo{author}{\bibfnamefont{D.~A.} \bibnamefont{Braje}},
  \bibinfo{title}{{Photoluminescence Decomposition Analysis : A Technique to
  Characterize N- V Creation in Diamond}}, \bibinfo{journal}{Phys. Rev. Appl.}
  \textbf{\bibinfo{volume}{12}}, \bibinfo{number}{044003}
  (\bibinfo{year}{2019}).

\bibitem[{\citenamefont{Bauch et~al.}(2020)\citenamefont{Bauch, Singh, Lee,
  Hart, Schloss, Turner, Barry, Pham, Bar-gill, Yelin et~al.}}]{Bauch2020}
\bibinfo{author}{\bibfnamefont{E.}~\bibnamefont{Bauch}},
  \bibinfo{author}{\bibfnamefont{S.}~\bibnamefont{Singh}},
  \bibinfo{author}{\bibfnamefont{J.}~\bibnamefont{Lee}},
  \bibinfo{author}{\bibfnamefont{C.~A.} \bibnamefont{Hart}},
  \bibinfo{author}{\bibfnamefont{J.~M.} \bibnamefont{Schloss}},
  \bibinfo{author}{\bibfnamefont{M.~J.} \bibnamefont{Turner}},
  \bibinfo{author}{\bibfnamefont{J.~F.} \bibnamefont{Barry}},
  \bibinfo{author}{\bibfnamefont{L.}~\bibnamefont{Pham}},
  \bibinfo{author}{\bibfnamefont{N.}~\bibnamefont{Bar-gill}},
  \bibinfo{author}{\bibfnamefont{S.~F.} \bibnamefont{Yelin}},
  \bibnamefont{et~al.}, \bibinfo{title}{{Decoherence of ensembles of
  nitrogen-vacancy centers in diamond}}, \bibinfo{journal}{Phys. Rev. B}
  \textbf{\bibinfo{volume}{102}}, \bibinfo{number}{134210}
  (\bibinfo{year}{2020}).

\bibitem[{\citenamefont{Tetienne et~al.}(2018)\citenamefont{Tetienne, {De
  Gille}, Broadway, Teraji, Lillie, McCoey, Dontschuk, Hall, Stacey, Simpson
  et~al.}}]{Tetienne2018}
\bibinfo{author}{\bibfnamefont{J.~P.} \bibnamefont{Tetienne}},
  \bibinfo{author}{\bibfnamefont{R.~W.} \bibnamefont{{De Gille}}},
  \bibinfo{author}{\bibfnamefont{D.~A.} \bibnamefont{Broadway}},
  \bibinfo{author}{\bibfnamefont{T.}~\bibnamefont{Teraji}},
  \bibinfo{author}{\bibfnamefont{S.~E.} \bibnamefont{Lillie}},
  \bibinfo{author}{\bibfnamefont{J.~M.} \bibnamefont{McCoey}},
  \bibinfo{author}{\bibfnamefont{N.}~\bibnamefont{Dontschuk}},
  \bibinfo{author}{\bibfnamefont{L.~T.} \bibnamefont{Hall}},
  \bibinfo{author}{\bibfnamefont{A.}~\bibnamefont{Stacey}},
  \bibinfo{author}{\bibfnamefont{D.~A.} \bibnamefont{Simpson}},
  \bibnamefont{et~al.}, \bibinfo{title}{{Spin properties of dense near-surface
  ensembles of nitrogen-vacancy centers in diamond}}, \bibinfo{journal}{Phys.
  Rev. B} \textbf{\bibinfo{volume}{97}}, \bibinfo{number}{085402}
  (\bibinfo{year}{2018}).

\bibitem[{\citenamefont{Pham et~al.}(2016)\citenamefont{Pham, Devience, Casola,
  Lovchinsky, Sushkov, Bersin, Lee, Urbach, Cappellaro, Park
  et~al.}}]{Pham2016}
\bibinfo{author}{\bibfnamefont{L.~M.} \bibnamefont{Pham}},
  \bibinfo{author}{\bibfnamefont{S.~J.} \bibnamefont{Devience}},
  \bibinfo{author}{\bibfnamefont{F.}~\bibnamefont{Casola}},
  \bibinfo{author}{\bibfnamefont{I.}~\bibnamefont{Lovchinsky}},
  \bibinfo{author}{\bibfnamefont{A.~O.} \bibnamefont{Sushkov}},
  \bibinfo{author}{\bibfnamefont{E.}~\bibnamefont{Bersin}},
  \bibinfo{author}{\bibfnamefont{J.}~\bibnamefont{Lee}},
  \bibinfo{author}{\bibfnamefont{E.}~\bibnamefont{Urbach}},
  \bibinfo{author}{\bibfnamefont{P.}~\bibnamefont{Cappellaro}},
  \bibinfo{author}{\bibfnamefont{H.}~\bibnamefont{Park}}, \bibnamefont{et~al.},
  \bibinfo{title}{{NMR technique for determining the depth of shallow
  nitrogen-vacancy centers in diamond}}, \bibinfo{journal}{Phys. Rev. B}
  \textbf{\bibinfo{volume}{93}}, \bibinfo{number}{045425}
  (\bibinfo{year}{2016}).

\bibitem[{\citenamefont{Loretz et~al.}(2015)\citenamefont{Loretz, Boss,
  Rosskopf, Mamin, Rugar, and Degen}}]{Loretz2015}
\bibinfo{author}{\bibfnamefont{M.}~\bibnamefont{Loretz}},
  \bibinfo{author}{\bibfnamefont{J.~M.} \bibnamefont{Boss}},
  \bibinfo{author}{\bibfnamefont{T.}~\bibnamefont{Rosskopf}},
  \bibinfo{author}{\bibfnamefont{H.~J.} \bibnamefont{Mamin}},
  \bibinfo{author}{\bibfnamefont{D.}~\bibnamefont{Rugar}}, \bibnamefont{and}
  \bibinfo{author}{\bibfnamefont{C.~L.} \bibnamefont{Degen}},
  \bibinfo{title}{{Spurious harmonic response of multipulse quantum sensing
  sequences}}, \bibinfo{journal}{Phys. Rev. X} \textbf{\bibinfo{volume}{5}},
  \bibinfo{number}{021009} (\bibinfo{year}{2015}).

\bibitem[{\citenamefont{Staudacher et~al.}(2015)\citenamefont{Staudacher,
  Raatz, Pezzagna, Meijer, Reinhard, Wrachtrup, and Meriles}}]{Staudacher2015}
\bibinfo{author}{\bibfnamefont{T.}~\bibnamefont{Staudacher}},
  \bibinfo{author}{\bibfnamefont{N.}~\bibnamefont{Raatz}},
  \bibinfo{author}{\bibfnamefont{S.}~\bibnamefont{Pezzagna}},
  \bibinfo{author}{\bibfnamefont{J.}~\bibnamefont{Meijer}},
  \bibinfo{author}{\bibfnamefont{F.}~\bibnamefont{Reinhard}},
  \bibinfo{author}{\bibfnamefont{J.}~\bibnamefont{Wrachtrup}},
  \bibnamefont{and} \bibinfo{author}{\bibfnamefont{C.~A.}
  \bibnamefont{Meriles}}, \bibinfo{title}{{Probing molecular dynamics at the
  nanoscale via an individual paramagnetic centre}}, \bibinfo{journal}{Nat.
  Commun.} \textbf{\bibinfo{volume}{6}}, \bibinfo{number}{8527}
  (\bibinfo{year}{2015}).

\bibitem[{\citenamefont{Rondin et~al.}(2014)\citenamefont{Rondin, Tetienne,
  Hingant, Roch, Maletinsky, and Jacques}}]{Rondin2014}
\bibinfo{author}{\bibfnamefont{L.}~\bibnamefont{Rondin}},
  \bibinfo{author}{\bibfnamefont{J.~P.} \bibnamefont{Tetienne}},
  \bibinfo{author}{\bibfnamefont{T.}~\bibnamefont{Hingant}},
  \bibinfo{author}{\bibfnamefont{J.~F.} \bibnamefont{Roch}},
  \bibinfo{author}{\bibfnamefont{P.}~\bibnamefont{Maletinsky}},
  \bibnamefont{and} \bibinfo{author}{\bibfnamefont{V.}~\bibnamefont{Jacques}},
  \bibinfo{title}{{Magnetometry with nitrogen-vacancy defects in diamond}},
  \bibinfo{journal}{Reports Prog. Phys.} \textbf{\bibinfo{volume}{77}},
  \bibinfo{number}{056503} (\bibinfo{year}{2014}).

\bibitem[{\citenamefont{Barry et~al.}(2020)\citenamefont{Barry, Schloss, Bauch,
  Turner, Hart, Pham, and Walsworth}}]{Barry2020}
\bibinfo{author}{\bibfnamefont{J.~F.} \bibnamefont{Barry}},
  \bibinfo{author}{\bibfnamefont{J.~M.} \bibnamefont{Schloss}},
  \bibinfo{author}{\bibfnamefont{E.}~\bibnamefont{Bauch}},
  \bibinfo{author}{\bibfnamefont{M.~J.} \bibnamefont{Turner}},
  \bibinfo{author}{\bibfnamefont{C.~A.} \bibnamefont{Hart}},
  \bibinfo{author}{\bibfnamefont{L.~M.} \bibnamefont{Pham}}, \bibnamefont{and}
  \bibinfo{author}{\bibfnamefont{R.~L.} \bibnamefont{Walsworth}},
  \bibinfo{title}{{Sensitivity optimization for NV-diamond magnetometry}},
  \bibinfo{journal}{Rev. Mod. Phys.} \textbf{\bibinfo{volume}{92}}
  (\bibinfo{year}{2020}).

\bibitem[{\citenamefont{Taylor et~al.}(2008)\citenamefont{Taylor, Cappellaro,
  Childress, Jiang, Budker, Hemmer, Yacoby, Walsworth, and Lukin}}]{Taylor2008}
\bibinfo{author}{\bibfnamefont{J.~M.} \bibnamefont{Taylor}},
  \bibinfo{author}{\bibfnamefont{P.}~\bibnamefont{Cappellaro}},
  \bibinfo{author}{\bibfnamefont{L.}~\bibnamefont{Childress}},
  \bibinfo{author}{\bibfnamefont{L.}~\bibnamefont{Jiang}},
  \bibinfo{author}{\bibfnamefont{D.}~\bibnamefont{Budker}},
  \bibinfo{author}{\bibfnamefont{P.~R.} \bibnamefont{Hemmer}},
  \bibinfo{author}{\bibfnamefont{A.}~\bibnamefont{Yacoby}},
  \bibinfo{author}{\bibfnamefont{R.}~\bibnamefont{Walsworth}},
  \bibnamefont{and} \bibinfo{author}{\bibfnamefont{M.~D.} \bibnamefont{Lukin}},
  \bibinfo{title}{{High-sensitivity diamond magnetometer with nanoscale
  resolution}}, \bibinfo{journal}{Nat. Phys.} \textbf{\bibinfo{volume}{4}},
  \bibinfo{number}{10} (\bibinfo{year}{2008}).

\bibitem[{\citenamefont{L{\"{u}}hmann et~al.}(2019)\citenamefont{L{\"{u}}hmann,
  John, Wunderlich, Meijer, and Pezzagna}}]{Luhmann2019}
\bibinfo{author}{\bibfnamefont{T.}~\bibnamefont{L{\"{u}}hmann}},
  \bibinfo{author}{\bibfnamefont{R.}~\bibnamefont{John}},
  \bibinfo{author}{\bibfnamefont{R.}~\bibnamefont{Wunderlich}},
  \bibinfo{author}{\bibfnamefont{J.}~\bibnamefont{Meijer}}, \bibnamefont{and}
  \bibinfo{author}{\bibfnamefont{S.}~\bibnamefont{Pezzagna}},
  \bibinfo{title}{{Coulomb-driven single defect engineering for scalable qubits
  and spin sensors in diamond}}, \bibinfo{journal}{Nat. Commun.}
  \textbf{\bibinfo{volume}{10}}, \bibinfo{number}{5956} (\bibinfo{year}{2019}).

\bibitem[{\citenamefont{Pehrsson and Mercer}(2000)}]{Pehrsson2000}
\bibinfo{author}{\bibfnamefont{P.~E.} \bibnamefont{Pehrsson}} \bibnamefont{and}
  \bibinfo{author}{\bibfnamefont{T.~W.} \bibnamefont{Mercer}},
  \bibinfo{title}{{Oxidation of heated diamond C (100): H surfaces}},
  \bibinfo{journal}{Surf. Sci.} \textbf{\bibinfo{volume}{460}}
  (\bibinfo{year}{2000}).

\bibitem[{\citenamefont{Riedel et~al.}(2004)\citenamefont{Riedel, Ristein, and
  Ley}}]{Riedel2004}
\bibinfo{author}{\bibfnamefont{M.}~\bibnamefont{Riedel}},
  \bibinfo{author}{\bibfnamefont{J.}~\bibnamefont{Ristein}}, \bibnamefont{and}
  \bibinfo{author}{\bibfnamefont{L.}~\bibnamefont{Ley}},
  \bibinfo{title}{{Recovery of surface conductivity of H-terminated diamond
  after thermal annealing in vacuum}}, \bibinfo{journal}{Phys. Rev. B}
  \textbf{\bibinfo{volume}{69}}, \bibinfo{number}{125338}
  (\bibinfo{year}{2004}).

\bibitem[{\citenamefont{Kucsko et~al.}(2018)\citenamefont{Kucsko, Choi, Choi,
  Maurer, Zhou, Landig, Sumiya, Onoda, Isoya, Jelezko et~al.}}]{Kucsko2018}
\bibinfo{author}{\bibfnamefont{G.}~\bibnamefont{Kucsko}},
  \bibinfo{author}{\bibfnamefont{S.}~\bibnamefont{Choi}},
  \bibinfo{author}{\bibfnamefont{J.}~\bibnamefont{Choi}},
  \bibinfo{author}{\bibfnamefont{P.~C.} \bibnamefont{Maurer}},
  \bibinfo{author}{\bibfnamefont{H.}~\bibnamefont{Zhou}},
  \bibinfo{author}{\bibfnamefont{R.}~\bibnamefont{Landig}},
  \bibinfo{author}{\bibfnamefont{H.}~\bibnamefont{Sumiya}},
  \bibinfo{author}{\bibfnamefont{S.}~\bibnamefont{Onoda}},
  \bibinfo{author}{\bibfnamefont{J.}~\bibnamefont{Isoya}},
  \bibinfo{author}{\bibfnamefont{F.}~\bibnamefont{Jelezko}},
  \bibnamefont{et~al.}, \bibinfo{title}{{Critical Thermalization of a
  Disordered Dipolar Spin System in Diamond}}, \bibinfo{journal}{Phys. Rev.
  Lett.} \textbf{\bibinfo{volume}{121}}, \bibinfo{number}{023601}
  (\bibinfo{year}{2018}).

\end{thebibliography}
\end{document}